\documentclass[twocolumn,aps,showpacs,amsfonts,amsmath,amssymb,floatfix]
{revtex4}
\usepackage[dvips]{graphicx}

\newcommand{\be}{\begin{equation}}
\newcommand{\ee}{\end{equation}}
\newcommand{\affA}{%
     Quantum Information Technology Group, \\
     National Institute of Information and Communications Technology
     (NICT), \\
     4-2-1 Nukui-kitamachi, Koganei, Tokyo 184-8795, Japan}
\newcommand{\affB}{%
     CREST, Japan Science and Technology Agency, 
     1-9-9 Yaesu, Chuoh-ku, Tokyo 103-0028, Japan}
\newcommand{\affC}{%
     Quantum Information Science Group, National Institute of
     Informatics (NII),\\
     2-1-2 Hitotsubashi, Chiyoda, Tokyo 101-8430, Japan}
\newcommand{\affD}{%
     Quantum Information Theory Group, Zentrum f\"{u}r Moderne Optik, \\
     Universit\"{a}t Erlangen-N\"{u}rnberg, 91058 Erlangen, Germany}

\begin{document}
\title{Implementation of projective measurements with linear optics \\
and continuous photon counting}
\author{Masahiro Takeoka}
\author{Masahide Sasaki}%
\affiliation{\affA}%
\affiliation{\affB}%

\author{Peter van Loock}
\affiliation{\affC}%
\author{Norbert L\"{u}tkenhaus}%
\affiliation{\affD}%

\begin{abstract}

We investigate the possibility of implementing a given projection
measurement using linear optics and arbitrarily fast feedforward
based on the continuous detection of photons. In particular, we
systematically derive the so-called Dolinar scheme that achieves
the minimum error discrimination of binary coherent states.
Moreover, we show that the Dolinar-type approach can also be
applied to projection measurements in the regime of photonic-qubit
signals. Our results demonstrate that for implementing a
projection measurement with linear optics, in principle, unit
success probability may be approached even without the use of
expensive entangled auxiliary states, as they are needed in all
known (near-)deterministic linear-optics proposals.

\end{abstract}
\pacs{03.67.Hk, 03.65.Ta, 42.50.Dv}
\date{\today}
\maketitle

\section{Introduction}

The implementation of positive operator-valued measures (POVMs)
for photonic quantum state signals is important for a variety of
quantum information protocols, in particular, for quantum
communication schemes such as quantum teleportation
\cite{Teleportation}, quantum key distribution
\cite{QuantumCrypto}, and collective decoding in quantum channel
coding \cite{Hausladen96,HSW,Fujiwara03}. Unlike conventional
optical detection technologies, POVMs for optical quantum
information protocols generally include a projection onto
superposition states or entangled states. In order to implement
such measurements, normally a nonlinear interaction of the signal
states (described by a Hamiltonian at least cubic in the optical
mode operators \cite{LloydBraunstein}) is needed. At present,
however, these nonlinear processes are hard to realize on the
level of single photons.

One possibility for inducing a nonlinear element is to exploit the
effective nonlinearity associated with a measurement.
In particular, for photonic-qubit states, universal gating
operations and hence any POVM for these states can be realized
asymptotically by using linear optics, photon counting, highly
entangled auxiliary states of $n$ photons, and conditional
dynamics (feedforward). Here, conditional dynamics means the
successive application of linear transformations on the remaining
modes conditioned upon the detection of a subset of modes
\cite{KLM01}. In the special case of a projection measurement,
corresponding to the discrimination of an orthogonal set of
states, perfect distinguishability is achieved in the asymptotic
limit of large $n$. However, with current technology, it is hard
to generate the entangled auxiliary states even for modest $n$.
More recent investigations, therefore, have focused on the
question whether one can implement a given measurement, or more
generally a set of universal quantum gates, via cheaper and/or
finite resources. For example, by applying the cluster-state model
of quantum computation \cite{RaussendorfBriegel} to linear optics
\cite{Nielsen}, the cost of the extra entangled resources may be
significantly reduced \cite{BrowneRudolph,Nielsen}.

In this paper, we address the following question: is it possible
to implement a given projection measurement in the asymptotic
limit of infinitely many, arbitrarily fast conditional-dynamics
steps without using any entangled auxiliary states at all?
Thereby, the intermediate detections upon which the conditional
dynamics relies are not supposed to be finite either, but they
shall be arbitrarily weak. In other words, instead of using
arbitrarily expensive auxiliary states of arbitrarily many photons
and a finite number of finite measurements plus feedforward
\cite{KLM01}, we employ infinitely many steps of feedforward
alone, based upon arbitrarily weak measurements. In fact, a nice
example for the latter approach was already given by Dolinar
\cite{Dolinar73} in the field of
quantum communication and detection theory, namely,
for the minimum error discrimination of binary coherent signals.

As an extension of the conventional signal detection theory,
quantum detection theory has been studied to give an optimal
signal decision strategy for noncommutative quantum signals
\cite{Helstrom_QDET}. It is motivated by fundamental interest, but
it also aims at investigating ultimate performance of optical
communication systems where information is usually carried by
coherent-state signals. Now, in this context, the simplest
scenario would be a communication scheme that is based on
the discrimination of the binary phase-shift keyed (BPSK) coherent
signals $\{ |\alpha\rangle,\,|\!-\! \alpha\rangle \}$ (in the
following, simply called the binary coherent signals). The optimal
POVM that discriminates these nonorthogonal signal states with the
minimum average error is described by a projection onto the
orthonormal basis consisting of superposition states of
$|\alpha\rangle$ and $|\!-\!\alpha\rangle$. This scheme is
sometimes called the Helstrom measurement \cite{Helstrom_QDET}.
Kennedy \cite{Kennedy73} first showed a simple physical model that
achieves the near-optimal measurement: the binary signal is
displaced to $\{|2\alpha\rangle,\,|0\rangle\}$ and then measured
by a photodetector that discriminates whether the signal contains
photons or not. Dolinar \cite{Dolinar73} (see also
\cite{Helstrom_QDET}) extended the Kennedy scheme demonstrating
that the perfect implementation of the Helstrom measurement is
possible by using linear optics, photon counting, and {\it
infinitesimally fast} feedforward. Other related proposals on the
implementation of the (near-)optimal measurement of binary
coherent states were given in
Refs.\cite{Bondurant93,Sasaki96_1,Sasaki96_2}.

\begin{figure}
\begin{center}
\includegraphics[width=80mm]{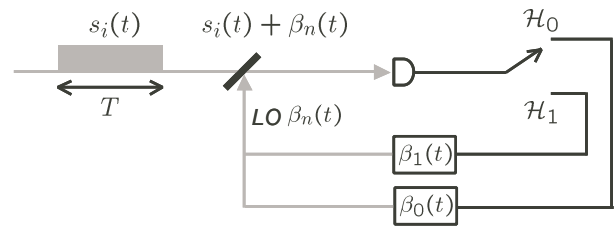}   %
\caption{\label{fig:Dolinar}The Dolinar receiver. The signal
$s_i(t)$ and the local oscillator are combined at a highly
transmissive beamsplitter such that the signal is displaced to
$s_i(t)+\beta_n(t)$. The appropriate choice of $\beta_n(t)$ at
$t'$ is determined by the number of the detected photons during
$[0, t')$ (For details, see the text). }
\end{center}
\end{figure}

A brief description of Dolinar's original proposal is the
following \cite{Dolinar73}. As shown in Fig.~\ref{fig:Dolinar},
the coherent signal field $s_i(t)$ ($i=0,1$) within the time
interval $T$ is displaced by one of the two local oscillators
(LOs) $\beta_0(t)$ and $\beta_1(t)$ and then incident into a
photodetector. The photodetector is assumed to have infinitesimal
time resolution and the functions $\beta_0(t)$ and $\beta_1(t)$
are appropriately chosen to minimize the error probability. The
choice of $\beta_0(t)$ or $\beta_1(t)$ at time $t'$ ($0\le t' <T$)
depends on whether the totally detected number of photons during
$[0,t')$ is even or odd. Therefore, once a signal photon triggers
a detector click, the current LO immediately has to be switched to
the other setting. After detecting the whole signal, one can infer
whether the signal was $s_0(t)$ or $s_1(t)$ by looking at the parity
of the number of the detected photons.
This scheme achieves the minimum error probability
after making an optimization of the likelyhood ratio of the binary signal
with optimal control theory \cite{Dolinar73,Geremia04}.
In the original paper \cite{Dolinar73}, the analysis was semiclassical
and later a fully quantum mechanical description was given \cite{Holevo82}.
More recently, the system performance under the realistic situation
including delay of feedforward, finite bandwidth and imperfect detection
has been investigated \cite{Yamazaki91,Geremia04}.

Compared to previous works on the Dolinar scheme, our contribution
here contains basically four new aspects. First, we revisit the
derivation of a Dolinar-type measurement scheme. However, instead
of focusing on the original signals $\{
|\alpha\rangle,\,|-\alpha\rangle \}$, we simply consider the
projection measurement onto the basis $\{
|\omega_0\rangle,\,|\omega_1\rangle \}$ that corresponds to the
minimum error discrimination of the signal states. We can then ask
whether one can discriminate the orthogonal states $\{
|\omega_0\rangle,\,|\omega_1\rangle \}$ via linear optics and
infinitely weak detections. In such a scheme, during the entire
measurement, the conditional states in every detection and
feedforward step must remain orthogonal. Using this constraint,
the derivation of the Dolinar protocol becomes simpler and more
transparent. Further, no complicated optimization procedures are
needed. In order to discuss the Dolinar receiver in the context of
linear-optics quantum information processing, we translate
Dolinar's original scheme from the time domain to the spatial
domain. In other words, infinitesimally fast feedforward is
replaced by an infinite use of spatial resources.

Secondly, we prove that there is no finite detection scheme that
attains the Helstrom bound for the minimum error discrimination of
binary coherent states. Such a no-go statement can be made by
using a set of criteria for the {\it exact} discrimination of
orthogonal states in a projection measurement \cite{vanLoock03}.
These criteria express in a simple way the requirement that the
orthogonal states must remain orthogonal after a linear-optics
transformation followed by the detection of a first mode. For
instance, in a Bell measurement for polarization-encoded photonic
qubits, this requirement can never be met and hence the Bell
measurement cannot be implemented with linear optics including
photon counting, finite steps of conditional dynamics, and
arbitrary auxiliary photon states
\cite{Luetkenhaus99,Calsamiglia01,vanLoock03}. Asymptotic schemes
\cite{KLM01} are not included in this approach. For the example of
binary coherent states investigated here, finite feedforward means
that the linear-optics transformation of the signal and auxiliary
modes before the first detection step involves a nonzero mixing
between the signal mode and the mode to be detected. Of course, in
the case of zero mixing, the orthogonality can be trivially
preserved. Here we will show how to accomplish the Helstrom
measurement based on feedforward using infinitely weak detections.

A third new aspect of our work is to demonstrate that the Dolinar
approach can be applied not only to the Helstrom measurement of
binary coherent states, but also to other measurements. In
particular, we apply the Dolinar approach to a projection
measurement in the regime of photonic-qubit signals. For a
particular example, for which any finite linear-optics scheme must
fail, we show that the Dolinar approach succeeds.

Last but not least, our analysis of the Dolinar scheme in the view
of the recent developments in linear-optics quantum information
processing demonstrates that by using infinitely many steps of
feedforward instead of arbitrarily expensive entangled photon
states, an asymptotically perfect efficiency of a projection
measurement is possible. Related to this, we note that there is
another proposal for an asymptotic linear-optics implementation of
a quantum measurement in which suboptimal unambiguous state
discrimination (USD) of $N$ symmetric coherent states for $N \ge
3$ is achieved without expensive auxiliary resources
\cite{vanEnk02}. For $N=2$, the optimal USD can be easily done
non-asymptotically by using a $50/50$ beamsplitter and photon
counting \cite{Huttner95}.

\section{Continuous Measurement via Linear Optics and Conditional Displacements}
In this section, we describe a linear-optics circuit in which a
sequence of arbitrarily weak measurements are asymptotically
combined to a continuous measurement. In particular, we discuss
the linear-optics circuit that corresponds to the Dolinar receiver
in the spatial domain. A fully quantum-mechanical description of
the Dolinar receiver was described in Ref.~\cite{Holevo82} by
applying a continuous photon-counting measurement based on the
quantum Markov process model \cite{Srinivas81,Srinivas82}.

\begin{figure}
\begin{center}
\includegraphics[width=80mm]{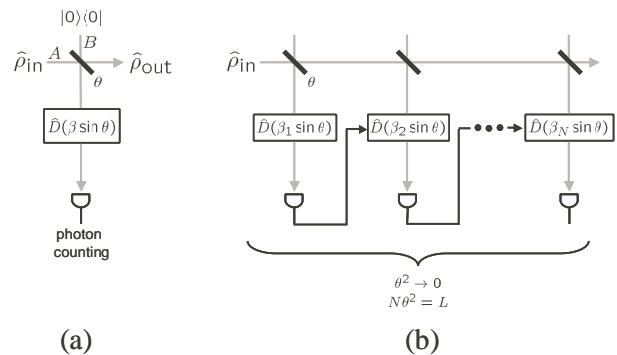}   %
\caption{\label{fig:ContinuousMeasurement} Continuous photon
counting with displacement operations. (a) Photodetection with a
displacement operation. The beamsplitter is parametrized by
$\theta$. (b) A sequential photodetection with displacement
operations in the limit of small $\theta$ (small reflectance) and
a large number of measurement steps. }
\end{center}
\end{figure}

In the following, we describe the continuous measurement based on
photon counting with a local oscillator by a sequence of linear
optics and photodetection. The equivalence between continuous
photon counting process and a sequence of beamsplitters and
photodetectors was firstly pointed out in Ref.~\cite{Ban94}. A
schematic of our measurement model is shown in
Fig.~\ref{fig:ContinuousMeasurement}(b). The scheme consists of a
sequence of weak photodetection steps. In each measurement step, a
small fraction of the signal state is reflected by a beamsplitter
and measured by a photodetector after a displacement operation
[Fig.~\ref{fig:ContinuousMeasurement}(a)]. The photodetector
counts the photon number of the signal.

After one measurement step, the output state conditioned upon the
number of detected photons $k$ is described by
\begin{equation}
\label{eq:in-out}
\hat{\rho}_{\rm out}^{(k)} =
\frac{ {\rm Tr_{\it B}}\, \left[
( \hat{I}^A \otimes \hat{\Pi}_k^B ) \, \hat{\rho}'_{\rm out} \right]
}{
{\rm Tr_{\it A,B}}\, \left[
( \hat{I}^A \otimes \hat{\Pi}_k^B ) \, \hat{\rho}'_{\rm out} \right] } ,
\end{equation}
where
\begin{eqnarray}
\hat{\rho}'_{\rm out} & = &
\hat{D}^B (\beta \sin\theta) 
\hat{B}^{AB} (\theta)
\left( \hat{\rho}_{\rm in}^A \otimes |0^B\rangle\langle0^B|
\right)
\nonumber\\
& & \times \hat{B}^{\dagger AB} (\theta)
\hat{D}^{\dagger B} (\beta \sin\theta) ,
\end{eqnarray}
and $\hat{\rho}_{\rm in}$ is the input state. Tr$_A$, Tr$_{A,B}$
denote the trace operations over the mode $A$ and the modes $A$
and $B$, respectively. $\hat{B}^{AB} (\phi)= \exp[
\phi(\hat{a}^{\dagger} \hat{b} - \hat{a} \hat{b}^{\dagger}) ]$ and
$\hat{D}^B (\alpha)= \exp( \alpha \hat{b}^{\dagger} - \alpha^*
\hat{b} )$ are the operators for a beamsplitter and a displacement
of mode $B$, where $\{ \hat{a},\,\hat{a}^{\dagger} \}$ and $\{
\hat{b},\,\hat{b}^{\dagger} \}$ are the annihilation and creation
operators for modes $A$ and $B$, respectively. Although
photodetection is described by a set of projection operators on
the Fock basis $\{ \hat{\Pi}_k = |k \rangle\langle k| \}$, we
assume that the parameter $\theta$ is sufficiently small such that
the probabilities of detecting more than two photons are
negligible. The complex number $\beta$ in the displacement
operator will be appropriately determined later. As it is well
known, this displacement operation can be realized using a
beamsplitter $\hat{B}(\theta)$ and a local oscillator of amplitude
$\beta$. For a pure state input $\hat{\rho}_{\rm in} = |\psi_{\rm
in}\rangle\langle\psi_{\rm in}|$, $\hat{\rho}'_{\rm
out}=|\psi'_{\rm out}\rangle\langle\psi'_{\rm out}|$ is given by
\begin{eqnarray}
\label{eq:out_dash1} |\psi'_{\rm out}\rangle & = & \hat{D}^B
(\beta \sin\theta) \hat{B}^{AB} (\theta) |\psi_{\rm in}^A\rangle
|0^B\rangle
\\
& = & e^{-\frac{1}{2}|\beta|^2 \sin^2\theta}
e^{\hat{b}^{\dagger}\beta \sin\theta} e^{- \hat{b}\,\beta^*
\sin\theta}
e^{-\hat{a}\hat{b}^{\dagger}\tan\theta} \nonumber\\
&& \times \, e^{\hat{a}^{\dagger} \hat{a}\ln\cos\theta}
e^{-\hat{b}^{\dagger} \hat{b}\ln\cos\theta}
e^{\hat{a}^{\dagger}\hat{b}\tan\theta} |\psi_{\rm in}^A\rangle
|0^B\rangle ,\nonumber
\end{eqnarray}
where we applied the Baker-Campbell-Hausdorff (BCH) formula and
for $\hat{B}^{AB}(\theta)$ an anti-normally ordered decomposition
formula such that $e^{\theta (\hat{a}^{\dagger} \hat{b} - \hat{a}
\hat{b}^{\dagger})} = e^{-\hat{a}\hat{b}^{\dagger}\tan\theta}
e^{\ln\cos\theta \, (\hat{a}^{\dagger}\hat{a} -
\hat{b}^{\dagger}\hat{b})}
e^{\hat{a}^{\dagger}\hat{b}\tan\theta}$. Throughout, we use the
notation $|\psi_{\rm in}^A\rangle|0^B\rangle \equiv |\psi_{\rm
in}^A\rangle\otimes|0^B\rangle$ etc.

Equation~(\ref{eq:out_dash1}) can be further simplified to
\begin{eqnarray}
\label{eq:out_dash} |\psi'_{\rm out}\rangle & = &
e^{-\frac{1}{2}|\beta|^2 \sin^2\theta} e^{(\beta \sin\theta -
\hat{a}\, \tan\theta) \hat{b}^{\dagger}} \nonumber\\ & & \times \,
e^{\hat{a}\, \beta^*  \frac{\sin^2\theta}{\cos\theta} }
e^{\hat{a}^{\dagger} \hat{a} \ln\cos\theta} |\psi_{\rm
in}^A\rangle|0^B\rangle .
\end{eqnarray}
Here we applied again the BCH formula such that $e^{-
\hat{b}\,\beta^* \sin\theta}
e^{-\hat{a}\hat{b}^{\dagger}\tan\theta} =
e^{-\hat{a}\hat{b}^{\dagger}\tan\theta} e^{\hat{a}\, \beta^*
\frac{\sin^2\theta}{\cos\theta} } e^{- \hat{b}\,\beta^*
\sin\theta}$, and we used $\hat{b}^{\dagger}\hat{b}\,|0^B\rangle =
\hat{b}\,|0^B\rangle = 0$.

Let us assume that the photodetector detects at most one photon.
We will now consider the two possible outcomes, i.e.
$\hat{\rho}'_{\rm out}$ is projected on $\hat{\Pi}_0$ or
$\hat{\Pi}_1$. Since these projectors are rank one operators, the
conditional output for each operator can be described by
\begin{equation}
\label{eq:two_outcomes}
\hat{\rho}_{\rm out}^{(k)} = \frac{
\hat{M}^{(k)} \hat{\rho}_{\rm in} \hat{M}^{(k)\dagger} }{
{\rm Tr_A}\, [
\hat{M}^{(k)} \hat{\rho}_{\rm in} \hat{M}^{(k)\dagger} ] }
\quad (k=0,1) ,
\end{equation}
where $\hat{M}^{(k)}$ is a Kraus operator for the $k$-photon
detection event. When no photon is detected at the photodetector,
the conditional output is given by
\begin{equation}
\label{eq:no-count-process} \hat{M}^{(0)} |\psi_{\rm in}\rangle =
e^{- \frac{1}{2} |\beta|^2 \sin^2\theta} e^{\hat{a}\, \beta^*
\frac{\sin^2\theta}{\cos\theta} } e^{\hat{a}^{\dagger} \hat{a} \ln
\cos \theta } |\psi_{\rm in}\rangle .
\end{equation}
Here we consider $N$ successive operations of
Eq.~(\ref{eq:no-count-process}) with the displacement
$\hat{D}(\beta_n \sin\theta)$ representing the $n$-th operation.
For example, for $N=2$, we have
\begin{widetext}
\begin{eqnarray}
\label{eq:twooperations} \hat{M}^{(0)}_2 \hat{M}^{(0)}_1
|\psi_{\rm in}\rangle
 =  \exp \left[- \frac{1}{2} (|\beta_1|^2 + |\beta_2|^2)
 \sin^2\theta \right]
\exp \left[\hat{a}\, \beta_2^* \frac{\sin^2\theta}{\cos\theta}
\right] \exp \left[\hat{a}\, \beta_1^*
\frac{\sin^2\theta}{\cos\theta} \,e^{-\ln \cos \theta}\right] \exp
\left[\hat{a}^{\dagger} \hat{a} \ln \cos^2 \theta \right]
|\psi_{\rm in}\rangle .\nonumber\\
\end{eqnarray}
Here we inserted the identity $e^{-\hat{a}^{\dagger} \hat{a} \ln
\cos \theta }e^{\hat{a}^{\dagger} \hat{a} \ln \cos \theta }$ and
then used $e^{\phi\hat{a}^{\dagger} \hat{a}} \,\hat{a}\,
e^{-\phi\hat{a}^{\dagger} \hat{a}}=\hat{a} e^{-\phi}$. For $N$
successive operations, the output is then given by
\begin{eqnarray}
\label{eq:N_times_no_count} \hat{M}^{(0)}_N \hat{M}^{(0)}_{N-1}
\cdots \hat{M}^{(0)}_1 |\psi_{\rm in}\rangle = \exp \left[
-\frac{1}{2} \sum^N_{n=1} |\beta_n|^2 \sin^2\theta \right] \exp
\left[ \hat{a} \sum^N_{n=1} \beta_n^* \cos^n \theta
\frac{\sin^2\theta}{\cos^{N+1} \theta} \right] \exp \left[
\hat{a}^{\dagger} \hat{a} \ln \cos^N \theta \right] |\psi_{\rm
in}\rangle .
\end{eqnarray}
Let us define the constant parameter
$L=N \theta^2$ and take the limit $\theta^2 \to 0$ and
$N \to \infty$. The displacement operations may now be described
in a continuous way via
\begin{eqnarray}
\label{eq:continuous_displacement}
& & \sum^N_{n=1} |\beta_n|^2 \sin^2\theta \approx
\sum^N_{n=1} |\beta_n|^2 \frac{L}{N} \to
\int^L_0 {\rm d}l \, |\beta (l)|^2 ,
\nonumber\\ & &
\sum^N_{n=1} \beta^*_n \cos^n \theta  \frac{\sin^2\theta}{\cos\theta} \approx
\sum^N_{n=1} \beta^*_n (\cos^N \theta)^{\frac{n}{N}} \theta^2 =
\sum^N_{n=1} \beta^*_n e^{-\frac{nL}{2N}} \frac{L}{N} \to
\int^L_0 {\rm d}l \, \beta^* (l) e^{-\frac{l}{2}} .
\end{eqnarray}
Using these relations, we obtain the output state
\begin{eqnarray}
\label{eq:0_L_no_count}
\hat{M}^{(0)}_N \hat{M}^{(0)}_{N-1} \cdots \hat{M}^{(0)}_1
|\psi_{\rm in}\rangle & = &
\exp \left[ -\frac{1}{2} \int^L_0 {\rm d}l \, |\beta(l)|^2 \right]
\exp \left[ \hat{a} e^{\frac{L}{2}}
\int^L_0 {\rm d}l \, \beta^*(l) e^{-\frac{l}{2}} \right]
\exp \left[ -\frac{L}{2} \hat{a}^{\dagger} \hat{a} \right]
|\psi_{\rm in}\rangle .
\end{eqnarray}
More generally, when no photon is detected from the $L_0$-th
detector to the $L_1$-th detector, we obtain the output state
\begin{eqnarray}
\label{eq:output_S_L1-L0}
\hat{\rho}_{\rm out} = \frac{
\hat{M}^{(0)}_{L_1} \hat{M}^{(0)}_{L_1-1} \cdots \hat{M}^{(0)}_{L_0}
\hat{\rho}_{\rm in}
\hat{M}^{(0)\dagger}_{L_0} \cdots
\hat{M}^{(0)\dagger}_{L_1-1} \hat{M}^{(0)\dagger}_{L_1} }{
{\rm Tr}\, [
\hat{M}^{(0)}_{L_1} \hat{M}^{(0)}_{L_1-1} \cdots \hat{M}^{(0)}_{L_0}
\hat{\rho}_{\rm in}
\hat{M}^{(0)\dagger}_{L_0} \cdots
\hat{M}^{(0)\dagger}_{L_1-1} \hat{M}^{(0)\dagger}_{L_1} ] }
= \frac{
\hat{S}_{L_1-L_0} \hat{\rho}_{\rm in} \hat{S}_{L_1-L_0}^{\dagger} }{
{\rm Tr}\, [
\hat{S}_{L_1-L_0} \hat{\rho}_{\rm in} \hat{S}_{L_1-L_0}^{\dagger} ] },
\end{eqnarray}
where
\begin{equation}
\label{eq:S_L1-L0}
\hat{S}_{L_1-L_0} =
\exp \left[ \hat{a} e^{\frac{L_1-L_0}{2}}
\int^{L_1}_{L_0} {\rm d}l \, \beta^*(l) e^{-\frac{l}{2}} \right]
\exp \left[ -\frac{L_1-L_0}{2} \hat{a}^{\dagger} \hat{a} \right] .
\end{equation}

However, when a photon is detected at the $L$-th photodetector,
the conditional operation is described by
\begin{equation}
\label{eq:count-process} \hat{M}^{(1)}_L |\psi_{\rm in}\rangle =
\exp \left[ -\frac{1}{2} |\beta_L|^2 \sin^2\theta \right] (
\beta_L \sin\theta - \hat{a} \tan\theta ) \exp \left[ \hat{a}
\beta_L^* \frac{\sin^2\theta}{\cos\theta} \right] \exp \left[
\hat{a}^{\dagger} \hat{a} \ln \cos \theta \right] |\psi_{\rm
in}\rangle .
\end{equation}
\end{widetext}
Upon taking the limit $\theta^2 \to 0$ and using the continuous
representation given in Eq.~(\ref{eq:continuous_displacement}),
all the exponential terms in Eq.~(\ref{eq:count-process}) approach
unity. The output now becomes
\begin{equation}
\label{eq:output_J_L}
\hat{\rho}_{\rm out} = \frac{
\hat{M}^{(1)}_L \hat{\rho}_{\rm in} \hat{M}^{(1)\dagger}_L }{
{\rm Tr}\, [
\hat{M}^{(1)}_L \hat{\rho}_{\rm in} \hat{M}^{(1)\dagger}_L ] }
= \frac{
\hat{J}_L \hat{\rho}_{\rm in} \hat{J}^{\dagger}_L }{
{\rm Tr}\, [
\hat{J}_L \hat{\rho}_{\rm in} \hat{J}^{\dagger}_L ] } ,
\end{equation}
where $ \hat{M}_L^{(1)} = \theta \hat{J}_L  $ and
\begin{equation}
\label{eq:J_L}
\hat{J}_L = \beta(L) - \hat{a} .
\end{equation}

Using these expressions, a continuous measurement based on
beamsplitting, displacement operations, and photon counting is
described via the operator 
\begin{equation}
\label{eq:SJSJ_}
\hat{S}_{L_{n}-L_{n-1}} \hat{J}_{L_{n-1}} \hat{S}_{L_{n-1}-L_{n-2}}
\cdots \hat{J}_{L_1} \hat{S}_{L_1-0} .
\end{equation}

This expression can also be obtained 
as a solution of the master equation of the system and 
this kind of conditional dynamics is called quantum jump process 
in terms of quantum statistics theory \cite{QuantumNoise}. 
Although analytical solutions can be obtained in our case, 
it should be noted when the system has complicated quantum jump e.g. 
the measurement has continuous outcome, one effective approach is 
the stochastic unraveling of the master equation. 
One of the successful applications of this approach is shown in 
Ref.~\cite{Wiseman93}.

In the next section, we will show how a given projective
measurement can be implemented via the apparatus discussed here.
Finally, we note that when $L$ is replaced by a time parameter
$T$, our formulation is equivalent to the conventional time domain
continuous measurement model including displacement operations
\cite{Holevo82}.

\section{Projective measurements via continuous measurement}

Projective measurements represent an important special case among
the generalized measurements. In this section, we show that the
Dolinar receiver can be systematically derived from an
orthogonality condition, similar to that used for analyzing the
exact distinguishability of orthogonal states in a projection
measurement \cite{vanLoock03}. First, we will apply this approach
to the original Dolinar receiver, that is, the discrimination of
binary coherent states. In a further example, we
examine a binary projection measurement onto a photonic-qubit
basis, which otherwise cannot be implemented with linear optics
including photon counting, finite steps of conditional dynamics,
and arbitrary auxiliary photon states.

Before discussing particular examples, we briefly summarize our
approach. The problem of implementing a complete projection
measurement $\{ \Pi_i = |\pi_i\rangle\langle\pi_i| \}$ can be
regarded as the problem of an exact discrimination of the
orthogonal signal states $|\pi_i\rangle$ \cite{vanLoock03}. In
order to achieve an exact discrimination, these signal states,
when conditionally transformed via partial measurements and
feedforward, must remain orthogonal after each step of the
intermediate measurements. In our approach, these measurements are
assumed to be arbitrarily weak, asymptotically corresponding to a
continuous measurement. Via the orthogonality constraint, we can
infer the input signal by counting the total number of detected
photons $N_{\rm tot}$. In the limit of infinitely many photodetection steps,
the final state must be in a vacuum state $|0\rangle$ due to the
energy loss at each step \cite{energyloss}.
Combining this fact with the condition
that the signals always remain mutually orthogonal, we know that
every possible result of the $N_{\rm tot}$-photon detection can be triggered
only by one of the two signal states with nonzero probability.
Eventually, we can infer the input signal state perfectly by
counting $N_{\rm tot}$.

\subsection{Minimum error discrimination
of binary coherent states}

The minimum error detection of the binary coherent states $\{
|\alpha\rangle, \, |-\alpha\rangle \}$ is achieved via the
projection operators corresponding to the orthogonal states
\begin{eqnarray}
\label{eq:Helstrom_measurement}
|\omega_0\rangle & = & \sqrt{\frac{1-P_e}{1-\kappa^2}} |\alpha\rangle
                 - \sqrt{\frac{P_e}{1-\kappa^2}} |\!-\!\alpha\rangle , \\
|\omega_1\rangle & = & \sqrt{\frac{P_e}{1-\kappa^2}} |\alpha\rangle
                 - \sqrt{\frac{1-P_e}{1-\kappa^2}} |\!-\!\alpha\rangle ,
\end{eqnarray}
where $\kappa=|\langle\alpha| \! - \! \alpha\rangle|$ and
\begin{equation}
\label{eq:min_err_probability}
P_e = \frac{1}{2} \left( 1-\sqrt{1-\kappa^2} \right) ,
\end{equation}
is the minimum error probability. For the sake of simplicity, we
assume that the a priori probabilities for the signals are equal.

Let us first prove that the projection onto
$\{|\omega_0\rangle,\,|\omega_1\rangle\}$ cannot be implemented
using linear optics, finite steps of conditional
dynamics, and arbitrary auxiliary states.
As for the detection mechanisms, we may restrict ourselves
to photon counting since homodyne detection with linear optics
never leads to non-Gaussian operation,
as required for our projection measurement.
Now, using the criteria for exact discrimination of orthogonal states
\cite{vanLoock03}, one finds that already the detection of a first
output mode, after mixing the signal mode with the auxiliary
modes, inevitably destroys the orthogonality. Defining an
arbitrary auxiliary state $|A\rangle$ with arbitrarily many modes,
the necessary conditions for preserving the orthogonality after
such a first detection (and hence for potentially enabling one to
exactly discriminate the states via further detections and
conditional transformations) are \cite{vanLoock03}
\begin{eqnarray}\label{criteria}
\langle A|\langle\omega_0| (\hat c^\dagger)^n\hat c^n
|\omega_1\rangle|A\rangle = 0,\quad \forall n=0,1,2,...\nonumber\\
\end{eqnarray}
Here, the 0th order ($n=0$) just corresponds to the orthogonality
of the signal states.
The annihilation operator
$\hat c$ represents the first mode being detected after the
linear-optics transformation. This output mode may be decomposed
into two parts of which one refers to the signal mode and the
other one to the auxiliary modes,
\begin{eqnarray}\label{decompose}
\hat c = \nu_1\hat a_1 + b_{\rm aux} \hat c_{\rm aux} + \gamma\,.
\end{eqnarray}
Here, $\nu_1\equiv U_{j1}$ is the complex entry of the unitary
matrix $U$ for describing the linear-optics mixing of the mode $j$
to be detected with the signal mode $1$. The mixing of mode $j$
with the auxiliary modes due to linear optics is described by the
annihilation operator $\hat c_{\rm aux}$, where $b_{\rm aux}$ is a
real parameter. The complex parameter $\gamma$ enables us to
include the possibility of phase-space displacements before the
detection \cite{vanLoock04inprep}. Note that without including
displacements and for signal states with a fixed number of
photons, arbitrary auxiliary states $|A\rangle$ cannot help to
provide nontrivial solutions to the conditions in
Eq.~(\ref{criteria}), if there are only trivial solutions without
an extra state $|A\rangle$ \cite{vanLoock03}. For the projection
onto $\{|\omega_0\rangle,\,|\omega_1\rangle\}$, however, the
signal states have an unfixed photon number. Thus, using an
auxiliary state $|A\rangle$ having unfixed number too
\cite{vanLoock03}, or employing a nonzero phase-space displacement
$\gamma\neq 0$ may indeed help.

After some algebra, for the first-order condition $n=1$ from
Eq.~(\ref{criteria}) using Eq.~(\ref{decompose}), we obtain
\begin{equation}\label{finalcondition}
|\nu_1|^2\,|\alpha|^2 = i \sqrt{1-\kappa^2}\,{\rm Im}\,\delta\,,
\end{equation}
where $\alpha\neq 0$ is the complex amplitude given by the signal
states, the parameter $\kappa=|\langle\alpha| \! - \!
\alpha\rangle|$ is also defined through the signal states, and
${\rm Im}\,\delta$ is the imaginary part of
\begin{equation}
\delta\equiv \nu_1^*\alpha^* (\gamma + \langle A|b_{\rm aux} \hat
c_{\rm aux}|A\rangle)\,.
\end{equation}
Since $\kappa^2 < 1$ for $\alpha\neq 0$, one can easily see that
the only solution to the condition in Eq.~(\ref{finalcondition})
is trivial, $\nu_1 = 0$. In other words, only if there is no
mixing at all between the mode $j$ to be detected and the signal
mode $1$, the orthogonality is (trivially) preserved. For any
finite mixing between the modes $j$ and $1$, exact discrimination
of the signal states is no longer possible. In our approach based
upon continuous measurement, every single conditional-dynamics
step is supposed to be arbitrarily weak corresponding to the limit
$\nu_1\to 0$. Thus, in this limit, the orthogonality condition may
be satisfied. However, typically, such a scheme does not provide
any information about the input signal states. Yet in the
following, we demonstrate that by combining infinitely many
arbitrarily weak detections of the signal mode, corresponding to a
continuous measurement of the signal mode, eventually perfect
state discrimination can be accomplished.

Suppose that the states $\{|\omega_0\rangle,\,|\omega_1\rangle\}$
are sent into the continuous measurement apparatus discussed in
the preceding section. When no photon is counted during
$[0,\,L_1)$, the signals evolve as
\begin{eqnarray}
\label{eq:no-count_Opt1}
\hat{S}_{L_1-0} |\omega_0\rangle & = &
\sqrt{\frac{1-P_e}{1-\kappa^2}} \,
\exp\left[ B \right]
\, |\alpha e^{-\frac{L_1}{2}}\rangle
\nonumber\\ & &
- \sqrt{\frac{P_e}{1-\kappa^2}} \,
\exp\left[ -B \right]
\, |\!-\!\alpha e^{-\frac{L_1}{2}}\rangle ,
\nonumber\\
\\
\label{eq:no-count_Opt2}
\hat{S}_{L_1-0} |\omega_1\rangle & = &
\sqrt{\frac{P_e}{1-\kappa^2}} \,
\exp\left[ B \right]
\, |\alpha e^{-\frac{L_1}{2}}\rangle
\nonumber\\ & &
- \sqrt{\frac{1-P_e}{1-\kappa^2}} \,
\exp\left[ -B \right]
\, |\!-\!\alpha e^{-\frac{L_1}{2}}\rangle ,
\nonumber\\
\end{eqnarray}
where
\begin{equation}
\label{eq:B}
B = \alpha \int^{L_1}_0 {\rm d}l \, \beta^*(l) e^{-\frac{l}{2}} .
\end{equation}
We note that these states are unnormalized. By calculating the
inner product between Eqs.~(\ref{eq:no-count_Opt1}) and
(\ref{eq:no-count_Opt2}), we can find the condition that these
signals are orthogonal,
\begin{eqnarray}
\label{eq:Dolinar_inner_product}
& & \sqrt{P_e(1-P_e)} \left\{
\exp\left[ B+B^* \right]
+\exp\left[ -(B+B^*) \right]
\right\} \nonumber\\
& & - \left\{ P_e \exp\left[ B-B^* \right]
+ (1-P_e) \exp\left[ -(B-B^*) \right] \right\}
\nonumber\\ & &
\times \exp \left[ -2 |\alpha|^2 e^{-L_1} \right] = 0 .
\end{eqnarray}
From the orthogonality condition
Eq.~(\ref{eq:Dolinar_inner_product}) with Eq.~(\ref{eq:B}), we
obtain the function for the displacement operation
\begin{equation}
\label{eq:LO}
\beta(l) = \pm \frac{ \alpha e^{-\frac{l}{2}} }{
\sqrt{ 1 - \exp[-4|\alpha|^2 (1-e^{-l})] } } .
\end{equation}
By defining the two solutions in Eq.~(\ref{eq:LO}) as
$\beta_+(l),\,\beta_-(l)$, respectively, we can derive the relations
\begin{eqnarray}
\label{eq:beta+}
\exp\left[ \alpha \int^{L_{j+1}}_{L_j} {\rm d}l \,
\beta_+(l) e^{-\frac{l}{2}}
\right] & = &
\frac{e^{-|\alpha|^2 e^{-L_{j+1}}}}{e^{-|\alpha|^2 e^{-L_{j}}}}
\frac{\sqrt{1-P_e^{L_{j+1}}}}{\sqrt{1-P_e^{L_{j}}}} ,
\nonumber\\
\end{eqnarray}
\begin{eqnarray}
\label{eq:beta-}
\exp\left[ \alpha \int^{L_{j+1}}_{L_j} {\rm d}l \,
\beta_-(l) e^{-\frac{l}{2}}
\right] & = &
\frac{e^{-|\alpha|^2 e^{-L_{j+1}}}}{e^{-|\alpha|^2 e^{-L_{j}}}}
\frac{\sqrt{P_e^{L_{j+1}}}}{\sqrt{P_e^{L_{j}}}} ,
\nonumber\\
\end{eqnarray}
where
\begin{equation}
P_e^{L} =
\frac{1}{2} \left( 1-\sqrt{1-\exp[-4|\alpha|^2 (1-e^{-L})]} \right) .
\end{equation}

In the following, we choose $\beta_+(l)$ as a displacement
function for $[0,\,L_1)$ ($\beta_-(l)$ leads to the same
conclusions) and assume that a photon is detected at the $L_1$-th
detector. The equations (\ref{eq:no-count_Opt1}),
(\ref{eq:no-count_Opt2}), (\ref{eq:beta+}), and (\ref{eq:beta-}) yield
\begin{eqnarray}
\label{eq:no-count_omega1(2)}
\hat{S}_{L_1-0} |\omega_0\rangle & \propto &
\sqrt{1-P_e} \frac{\sqrt{1-P_e^{L_1}}}{\sqrt{1-P_e^{0}}}
| \alpha e^{-\frac{L_1}{2}}\rangle
\nonumber\\
& &
- \sqrt{P_e} \frac{\sqrt{P_e^{L_1}}}{\sqrt{P_e^{0}}}
| \!-\! \alpha e^{-\frac{L_1}{2}}\rangle ,
\\
\label{eq:no-count_omega2(2)}
\hat{S}_{L_1-0} |\omega_1\rangle & \propto &
\sqrt{P_e} \frac{\sqrt{1-P_e^{L_1}}}{\sqrt{1-P_e^{0}}}
|\alpha e^{-\frac{L_1}{2}}\rangle
\nonumber\\
& &
- \sqrt{1-P_e} \frac{\sqrt{P_e^{L_1}}}{\sqrt{P_e^{0}}}
| \!-\! \alpha e^{-\frac{L_1}{2}}\rangle ,
\end{eqnarray}
where unimportant global coefficients are omitted.
As for the states after detecting a photon,
by using the relations
\begin{eqnarray}
\label{eq:J1}
\beta_+(L_1) - \alpha e^{-\frac{L_1}{2}} =
\frac{2 P_e^{L_1} \alpha e^{-\frac{L_1}{2} }}{
\sqrt{1-\exp[-4|\alpha|^2 (1-e^{-L_1})]}} ,
\nonumber\\
\end{eqnarray}
\begin{eqnarray}
\label{eq:J2}
\beta_+(L_1) + \alpha e^{-\frac{L_1}{2}} =
\frac{2 (1-P_e^{L_1}) \alpha e^{-\frac{L_1}{2} }}{
\sqrt{1-\exp[-4|\alpha|^2 (1-e^{-L_1})]}} ,
\nonumber\\
\end{eqnarray}
we find
\begin{eqnarray}
\label{eq:JS0}
\hat{J}_{L_1} \hat{S}_{L_1-0} |\omega_0\rangle
& \propto &
\sqrt{1-P_e} \frac{\sqrt{1-P_e^{L_1}}}{\sqrt{1-P_e^{0}}} P_e^{L_1}
             |\alpha e^{-\frac{L_1}{2}}\rangle
\nonumber\\
& &
- \sqrt{P_e} \frac{\sqrt{P_e^{L_1}}}{\sqrt{P_e^{0}}} (1-P_e^{L_1})
             |\!-\!\alpha e^{-\frac{L_1}{2}}\rangle
\nonumber\\
& \propto &
\sqrt{1-P_e} \frac{\sqrt{P_e^{L_1}}}{\sqrt{1-P_e^{0}}}
             |\alpha e^{-\frac{L_1}{2}}\rangle
\nonumber\\
& &
- \sqrt{P_e} \frac{\sqrt{1-P_e^{L_1}}}{\sqrt{P_e^{0}}}
             |\!-\!\alpha e^{-\frac{L_1}{2}}\rangle ,
\end{eqnarray}
and similarly,
\begin{eqnarray}
\label{eq:JS1}
\hat{J}_{L_1} \hat{S}_{L_1-0} |\omega_1\rangle
& \propto &
\sqrt{P_e} \frac{\sqrt{P_e^{L_1}}}{\sqrt{1-P_e^{0}}}
             |\alpha e^{-\frac{L_1}{2}}\rangle
\nonumber\\
& &
- \sqrt{1-P_e} \frac{\sqrt{1-P_e^{L_1}}}{\sqrt{P_e^{0}}}
             |\!-\!\alpha e^{-\frac{L_1}{2}}\rangle .
\nonumber\\
\end{eqnarray}
Comparing them with Eqs.~(\ref{eq:no-count_omega1(2)}) and
(\ref{eq:no-count_omega2(2)}), we can easily see that these states
are still orthogonal to each other.
Also, we can derive the relations,
\begin{eqnarray}
\label{eq:JS_a+_b+}
& &\hat{J}_{L_{j+1}} \hat{S}_{L_{j+1}-L_j}
|\alpha e^{-\frac{L_j}{2}} \rangle
\nonumber\\
& &= \frac{1}{2}e^{-4|\alpha|^2 (1-e^{-L_{j+1}})}
\frac{\sqrt{P_e^{L_{j+1}}}}{\sqrt{1-P_e^{L_j}}} |\alpha
e^{-\frac{L_{j+1}}{2}} \rangle ,
\\
& &\label{eq:JS_a-_b+}
\hat{J}_{L_{j+1}} \hat{S}_{L_{j+1}-L_j}
|\!-\!\alpha e^{-\frac{L_j}{2}} \rangle
\nonumber\\
& &= \frac{1}{2}e^{-4|\alpha|^2 (1-e^{-L_{j+1}})}
\frac{\sqrt{1-P_e^{L_{j+1}}}}{\sqrt{P_e^{L_j}}}
|\!-\!\alpha e^{-\frac{L_{j+1}}{2}} \rangle ,
\nonumber\\
\end{eqnarray}
for the measurement process
with the displacement function $\beta_+ (l)$, and
\begin{eqnarray}
\label{eq:JS_a+_b-}
& &\hat{J}_{L_{j+1}} \hat{S}_{L_{j+1}-L_j}
|\alpha e^{-\frac{L_j}{2}} \rangle
\\
& &= \frac{1}{2}e^{-4|\alpha|^2 (1-e^{-L_{j+1}})}
\frac{\sqrt{1-P_e^{L_{j+1}}}}{\sqrt{P_e^{L_j}}}
|\alpha e^{-\frac{L_{j+1}}{2}} \rangle ,
\nonumber\\
\label{eq:JS_a-_b-}
& &\hat{J}_{L_{j+1}} \hat{S}_{L_{j+1}-L_j} |\!-\!\alpha e^{-L_j} \rangle
\\
& &= \frac{1}{2}e^{-4|\alpha|^2 (1-e^{-L_{j+1}})}
\frac{\sqrt{P_e^{L_{j+1}}}}{\sqrt{1-P_e^{L_j}}} |\!-\!\alpha
e^{-\frac{L_{j+1}}{2}} \rangle , \nonumber
\end{eqnarray}
for the measurement process
with the displacement function $\beta_- (l)$,
where $(P_e^{L_{j+1}} (1-P_e^{L_{j+1}}))^{1/2}=
e^{-4|\alpha|^2 (1-e^{-L_{j+1}})}/2$

From Eqs. (\ref{eq:beta+}), (\ref{eq:beta-}), (\ref{eq:JS0}) and
(\ref{eq:JS1}), we can determine the local oscillator function
$\beta(l)$ for the next zero photon counting process
$\hat{S}_{L_2-L_1}$. In our case, the orthogonality condition
requires using $\beta_-(l)$ for $\hat{S}_{L_2-L_1}$, as the output
states $\hat{S}_{L_2-L_1}\hat{J}_{L_1}\hat{S}_{L_1-0}
|\omega_i\rangle$ ($i=0,1$) have the similar structure as in
Eqs.~(\ref{eq:JS0}) and (\ref{eq:JS1}), respectively, except that
$L_1$ is replaced by $L_2$. Eventually, one can preserve the
orthogonality by switching the sign of the displacement function
$\beta_{\pm}(l)$ each time a photon is detected. After applying
many of these operations, the signals evolve as
$\hat{S}_{L-L_{N_{\rm tot}}}
\hat{J}_{L_{N_{\rm tot}}}
\hat{S}_{L_{N_{\rm tot}}-L_{{N_{\rm tot}}-1}} \cdots \hat{J}_{L_1}
\hat{S}_{L_1-0} |\omega_i\rangle$, $(i=0,1)$, where $N_{\rm tot}$ is the
total number of detected photons $( N_{\rm tot}=0,1 \cdots ,\infty)$
\cite{finiteN_tot} and we will take an infinitely large $L$.
The final states are now derived with the help of
Eqs.~(\ref{eq:JS_a+_b+}), (\ref{eq:JS_a-_b+}), (\ref{eq:JS_a+_b-}),
and (\ref{eq:JS_a-_b-}),
and classified via the parity of $N_{\rm tot}$.
When $N_{\rm tot}$ is zero or even, the output state for the input
$|\omega_0\rangle$ is given by
\begin{widetext}
\begin{eqnarray}
\label{eq:omega_0_odd}
|\omega_0^L\rangle & = &
\hat{S}_{L-L_{N_{\rm tot}}} \hat{J}_{L_{N_{\rm tot}}}
\hat{S}_{L_{N_{\rm tot}}-L_{{N_{\rm tot}}-1}} \cdots
\hat{J}_{L_1} \hat{S}_{L_1-0}
|\omega_0\rangle
\nonumber\\ & \propto &
\sqrt{1-P_e} \frac{\sqrt{P_e^{L_1}}}{\sqrt{1-P_e^{0}}}
\frac{\sqrt{1-P_e^{L_2}}}{\sqrt{P_e^{L_1}}}
\cdots \frac{\sqrt{1-P_e^L}}{\sqrt{1-P_e^{L_{N_{\rm tot}}}}}
|\alpha e^{-\frac{L}{2}} \rangle
- \sqrt{P_e} \frac{\sqrt{1-P_e^{L_1}}}{\sqrt{P_e^{0}}}
\frac{\sqrt{P_e^{L_2}}}{\sqrt{1-P_e^{L_1}}}
\cdots \frac{\sqrt{P_e^L}}{\sqrt{P_e^{L_{N_{\rm tot}}}}}
|\!-\!\alpha e^{-\frac{L}{2}} \rangle
\nonumber\\ & \propto &
\sqrt{1-P_e} \sqrt{1-P_e^L} |\alpha e^{-\frac{L}{2}} \rangle
- \sqrt{P_e} \sqrt{P_e^L} |\!-\!\alpha e^{-\frac{L}{2}} \rangle
\nonumber\\ & = &
e^{ -\frac{1}{2} |\alpha|^2 e^{-L} }
\sum^{\infty}_{m=0} \frac{1}{\sqrt{m!}} \left\{
\begin{array}{ll}
  \left( \sqrt{1-P_e}\sqrt{1-P_e^L} - \sqrt{P_e}\sqrt{P_e^L} \right)
  \left( \alpha e^{-\frac{L}{2}} \right)^m |m\rangle &
  \qquad \mbox{$m$ is zero or even} \\
  \left( \sqrt{1-P_e}\sqrt{1-P_e^L} + \sqrt{P_e}\sqrt{P_e^L} \right)
  \left( \alpha e^{-\frac{L}{2}} \right)^m |m\rangle &
  \qquad \mbox{$m$ is odd}
\end{array}
\right. \nonumber\\ & = &
\left( \sqrt{1-e^{-4|\alpha|^2}} +
\frac{|\alpha|^2 e^{-4|\alpha|^2}}{ \sqrt{1-e^{-4|\alpha|^2}} } e^{-L}
\cdots \right) |0\rangle + (1+\cdots) \alpha e^{-\frac{L}{2}} |1\rangle
+ \frac{1}{\sqrt{2}} \left( \sqrt{1-e^{-4|\alpha|^2}} + \cdots \right)
\alpha^2 e^{-L} |2\rangle + \cdots, \nonumber\\
\end{eqnarray}
where $|m\rangle$ is the $m$-photon number state and note that
$P_e^0 = 1/2$ and $\lim_{L \to \infty} P_e^L = P_e$. In the last
line, the terms only up to the order of $e^{-L}$ are given.
Similarly, for the input $|\omega_1\rangle$, we obtain
\begin{eqnarray}
\label{eq:omega_1_odd}
|\omega_1^L\rangle & = &
\hat{S}_{L-L_{N_{\rm tot}}} \hat{J}_{L_{N_{\rm tot}}}
\hat{S}_{L_{N_{\rm tot}}-L_{{N_{\rm tot}}-1}} \cdots
\hat{J}_{L_1} \hat{S}_{L_1-0}
|\omega_1\rangle
\nonumber\\ & \propto &
\sqrt{P_e}\sqrt{1-P_e^L} | \alpha e^{-\frac{L}{2}} \rangle
- \sqrt{1-P_e}\sqrt{P_e^L} |\!-\!\alpha e^{-\frac{L}{2}} \rangle
\nonumber\\ & = &
e^{ -\frac{1}{2} |\alpha|^2 e^{-L} }
\sum^{\infty}_{m=0} \frac{1}{\sqrt{m!}} \left\{
\begin{array}{ll}
  \left( \sqrt{P_e}\sqrt{1-P_e^L} - \sqrt{1-P_e}\sqrt{P_e^L} \right)
  \left( \alpha e^{-\frac{L}{2}} \right)^m |m\rangle &
  \qquad \mbox{$m$ is zero or even} \\
  \left( \sqrt{P_e}\sqrt{1-P_e^L} + \sqrt{1-P_e}\sqrt{P_e^L} \right)
  \left( \alpha e^{-\frac{L}{2}} \right)^m |m\rangle &
  \qquad \mbox{$m$ is odd}
\end{array}
\right. \nonumber\\ & = &
\left(
- \frac{|\alpha|^2 e^{-2|\alpha|^2}}{\sqrt{1-e^{-4|\alpha|^2}}} e^{-L}
+ \cdots \right)
|0\rangle + \left( e^{-2|\alpha|^2} + \cdots \right)
\alpha e^{-\frac{L}{2}} |1\rangle + \cdots ,
\end{eqnarray}
\end{widetext}
where the last line also shows the terms up to the order of $e^{-L}$.
From Eqs.~(\ref{eq:omega_0_odd}) and (\ref{eq:omega_1_odd}),
we can check that these two possible outputs are still orthogonal
to each other and, even in the approximated form,
the orthogonality is satisfied in any order of $L$.
For example, the inner product between
$|\omega_0^L\rangle$ and $|\omega_1^L\rangle$ up to the order of $e^{-L}$
is computed as
\begin{eqnarray}
\label{eq:inner_product_approx}
\langle\omega_0^L|\omega_1^L\rangle & \approx &
- \sqrt{1-e^{-4|\alpha|^2}}
\frac{|\alpha|^2 e^{-2|\alpha|^2}}{ \sqrt{1-e^{-4|\alpha|^2}} } e^{-L}
\langle 0|0 \rangle
\nonumber\\ & &
+ |\alpha|^2 e^{-2|\alpha|^2} e^{-L} \langle 1|1 \rangle
\nonumber\\ & = & 0.
\end{eqnarray}

Then, upon taking the limit of $L \to \infty$ and normalizing
Eq.~(\ref{eq:omega_0_odd}), we know that $|\omega_0^L\rangle$
approaches the vacuum state. We also find from
Eq.~(\ref{eq:omega_1_odd}) that the state $|\omega_1^L\rangle$
converges to $|1\rangle$. However, since {\it all} the terms in
Eq.~(\ref{eq:omega_1_odd}) decrease exponentially for large $L$,
we know that $|\omega_1^L\rangle$ does not occur at all in the
limit of $L \to \infty$. Hence we have
\begin{eqnarray}
\label{eq:omega_0_odd_final}
|\omega_0^L\rangle
& \to & |0\rangle ,
\\
\label{eq:omega_1_odd_final}
|\omega_1^L\rangle
& \to & 0 ,
\end{eqnarray}
and thus, if the number of the totally detected photons $N_{\rm
tot}$ is zero or even, we can unambiguously identify the input
state $|\omega_0\rangle$.

On the other hand, when $N_{\rm tot}$ is odd, the final states are given by
\begin{eqnarray}
\label{eq:omega_0_even}
|\omega_0^L\rangle
& \to & 0 ,
\\
\label{eq:omega_1_even}
|\omega_1^L\rangle
& \to & |0\rangle ,
\end{eqnarray}
and hence, in this case, we can unambiguously infer the input
state $|\omega_1\rangle$. Let us finally emphasize that our model
may be also translated into the time domain by replacing the
spatial parameter $l$ by the time $t$. Therefore, our approach
provides an alternative derivation of the original Dolinar
receiver. This approach is relatively simple, because it only
relies upon the preservation of the orthogonality of the signals
during the entire measurement process.

\subsection{Photonic-qubit signals}

In this subsection, we show that our measurement model is not only
applicable to coherent-state signals, but it may also be applied
to other types of signals, for example, photon-number qubits. As
an example, let us consider the projection measurement onto the
basis $|\omega_{0,1}\rangle = ( |0\rangle \pm |1\rangle )/\sqrt{2}$,
for which one can derive a no-go statement in terms of the
criteria of Ref.\cite{vanLoock03}.

As for the exact discrimination of $|\omega_{0,1}\rangle$, let us
first briefly discuss the no-go statement for any linear-optics
scheme based on photon counting, finite steps of conditional
dynamics, and arbitrary auxiliary states. Since the signal states
have an unfixed number of photons, using an auxiliary state
$|A\rangle$ with unfixed photon number or employing phase-space
displacements may help to exactly discriminate the signal states
\cite{vanLoock03}. For the signal states $|\omega_{0,1}\rangle$, the
necessary conditions in any conditional-dynamics scheme become
\cite{vanLoock03}
\begin{eqnarray}\label{criteria2}
\langle A|\langle\omega_0| (\hat c^\dagger)^n\hat c^n
|\omega_1\rangle|A\rangle = 0,\quad \forall n=0,1,2,...
\end{eqnarray}
Here, we use the same definitions as in the preceding section. By
inserting the decomposition of Eq.~(\ref{decompose}) into the
first-order $n=1$ condition, we obtain now
\begin{equation}\label{finalcondition2}
|\nu_1|^2 = 2i \,{\rm Im}\,\delta\,,
\end{equation}
where now ${\rm Im}\,\delta$ is the imaginary part of
\begin{equation}
\delta\equiv \nu_1(\gamma^* + \langle A|b_{\rm aux} \hat c_{\rm
aux}^\dagger|A\rangle)\,.
\end{equation}
Again, only the trivial solution $\nu_1=0$ exists. Thus, there is
no linear-optics scheme based on a finite first detection step
that achieves the exact discrimination of $|\omega_{0,1}\rangle$. Let
us now see how a continuous-measurement based scheme leads to the
perfect discrimination of the signal states.

The design for the measurement apparatus is the same as
that of the preceding subsection. So we send the states
$\{ |\omega_0\rangle,\,|\omega_1\rangle \}$ into the measurement
apparatus. Let us assume that the first photon is detected at the
$L_1$-th detector and no photon is counted during $[0,\,L_1)$. The
outcoming signals are then given by
\begin{eqnarray}
\label{eq:qubit_no-count_1-0}
\hat{S}_{L_1-0} |\omega_0\rangle & = &
\left( 1 + \int^{L_1}_0 {\rm d}l \, \beta^*(l) e^{-\frac{l}{2}} \right)
|0\rangle + e^{-\frac{L_1}{2}} |1\rangle ,
\nonumber\\
\\
\label{eq:qubit_no-count_1-1}
\hat{S}_{L_1-0} |\omega_1\rangle & = &
\left( 1 - \int^{L_1}_0 {\rm d}l \, \beta^*(l) e^{-\frac{l}{2}} \right)
|0\rangle - e^{-\frac{L_1}{2}} |1\rangle .
\nonumber\\
\end{eqnarray}
The orthogonality condition for Eqs.~(\ref{eq:qubit_no-count_1-0})
and (\ref{eq:qubit_no-count_1-1}) implies an appropriate local
oscillator function for $\{ |\omega_0\rangle,\,|\omega_1\rangle \}$,
namely
\begin{equation}
\label{eq:qubit_LO_L_1-0}
\beta_{L_1-0} (l) = \pm \frac{ e^{-\frac{l}{2}} }{
2 \sqrt{ 1-e^{-l} } } .
\end{equation}
When choosing the plus solution for $\beta_{L_1-0} (l)$,
Eqs.~(\ref{eq:qubit_no-count_1-0}) and
(\ref{eq:qubit_no-count_1-1}) can be rewritten as
\begin{eqnarray}
\label{eq:qubit_no-count_2-0} \hat{S}_{L_1-0} |\omega_0\rangle & =
& \left( 1 + \sqrt{1-e^{-L_1}} \right) |0\rangle +
e^{-\frac{L_1}{2}} |1\rangle ,
\nonumber\\
\end{eqnarray}
\begin{eqnarray}
\label{eq:qubit_no-count_2-1} \hat{S}_{L_1-0} |\omega_1\rangle & =
& \left( 1 - \sqrt{1-e^{-L_1}} \right) |0\rangle -
e^{-\frac{L_1}{2}} |1\rangle .\nonumber\\
\end{eqnarray}

After detecting a photon at $L_1$, the states become
\begin{eqnarray}
\label{eq:qubit_JS0}
\hat{J}_{L_1} \hat{S}_{L_1-0} |\omega_0\rangle & \propto &
\left( 1 - \sqrt{1-e^{-L_1}} \right) |0\rangle
+ e^{-\frac{L_1}{2}} |1\rangle ,
\nonumber\\
\\
\label{eq:qubit_JS1}
\hat{J}_{L_1} \hat{S}_{L_1-0} |\omega_1\rangle & \propto &
\left( 1 + \sqrt{1-e^{-L_1}} \right) |0\rangle
- e^{-\frac{L_1}{2}} |1\rangle ,
\nonumber\\
\end{eqnarray}
and these are still orthogonal to each other.

Let us consider the next interval $(L_1,\,L_2)$, where we assume
that the second photon is detected at $L_2$. Defining the local
oscillator function $\beta_{L_2-L_1} (l)$, we obtain
\begin{widetext}
\begin{eqnarray}
\label{eq:qubit_SJS0}
\hat{S}_{L_2-L_1} \hat{J}_{L_1} \hat{S}_{L_1-0} |\omega_0\rangle & \propto &
\left( 1 - \sqrt{1-e^{-L_1}} + e^{-\frac{L_1}{2}}
\int^{L_2}_{L_1} {\rm d}l \, \beta_{L_2-L_1}^* (l) e^{-\frac{l}{2}}
\right) |0\rangle + e^{-\frac{L_2}{2}} |1\rangle ,
\\
\label{eq:qubit_SJS1}
\hat{S}_{L_2-L_1} \hat{J}_{L_1} \hat{S}_{L_1-0} |\omega_1\rangle & \propto &
\left( 1 + \sqrt{1-e^{-L_1}} - e^{-\frac{L_1}{2}}
\int^{L_2}_{L_1} {\rm d}l \, \beta_{L_2-L_1}^* (l) e^{-\frac{l}{2}}
\right) |0\rangle - e^{-\frac{L_2}{2}} |1\rangle .
\end{eqnarray}
\end{widetext}
Now the orthogonality condition implies
\begin{equation}
\label{eq:qubit_LO2}
\beta_{L_2-L_1} (l) = - \frac{ e^{-\frac{l-L_1}{2}} }{
2 \sqrt{1-e^{-l}} } .
\end{equation}
More generally, the local oscillator function for the interval
$(L_j ,\, L_{j+1})$ is given by
\begin{eqnarray}
\label{eq:qubit_LOgeneral} \beta_{L_{j+1} - L_j} (l) = (-1)^j
\frac{ e^{-\frac{l-L_j}{2}} }{ 2 \sqrt{1-e^{-l}} } , \qquad
(L_0=0) .\nonumber\\
\end{eqnarray}

Using Eq.~(\ref{eq:qubit_LOgeneral}), the signal states after the
whole detection process are described by $\hat{S}_{L-L_N}
\hat{J}_{L_{N_{\rm tot}}} \hat{S}_{L_{N_{\rm tot}}-L_{{N_{\rm tot}}-1}}
\cdots \hat{J}_{L_1}
\hat{S}_{L_1-0} |\omega_i\rangle$, $(i=0,1)$, where $N_{\rm tot}$ is the total
number of detected photons and $(L \to \infty,\, N_{\rm tot}=0,1 \cdots
,\infty)$. When $N_{\rm tot}$ is zero or even, the final states are
\begin{eqnarray}
\label{eq:qubit_odd}
|\omega_0^L\rangle & \propto &
\left( 1+\sqrt{1-e^{-L}} \right) |0\rangle
+ e^{-\frac{L}{2}} |1\rangle
\nonumber\\ & = &
\left( 2 - \frac{1}{2}e^{-L} - \cdots \right) |0\rangle
+ e^{-\frac{L}{2}} |1\rangle
\nonumber\\ & \to & |0\rangle ,
\\
|\omega_1^L\rangle & \propto &
\left( 1-\sqrt{1-e^{-L}} \right) |0\rangle
- e^{-\frac{L}{2}} |1\rangle
\nonumber\\ & = &
\left( \frac{1}{2} e^{-L} + \cdots \right) |0\rangle
- e^{-\frac{L}{2}} |1\rangle
\nonumber\\ & \to & 0 .
\end{eqnarray}
These are still orthogonal to each other in any order of $L$.
On the other hand, when $N_{\rm tot}$ is odd, we have
\begin{eqnarray}
\label{eq:qubit_odd}
|\omega_0^L\rangle & \to & 0 ,
\\
|\omega_1^L\rangle & \to & |0\rangle .
\end{eqnarray}
Eventually, one finds that it is possible to distinguish
$|\omega_0\rangle$ and $|\omega_1\rangle$ perfectly by checking the
parity of the totally detected photon number. Let us finally
mention that the projection onto arbitrary orthogonal
superpositions of the vacuum state $|0\rangle$ and the single
photon state $|1\rangle$,
\begin{eqnarray}
\label{eq:0pm1_basis}
|\omega_0\rangle & = &
f_0 |0\rangle + f_1 e^{i\varphi} |1\rangle,
\\
|\omega_1\rangle & = &
f_1 |0\rangle - f_0 e^{i\varphi} |1\rangle,
\end{eqnarray}
can be achieved in the same manner as described above.

\section{Conclusions}

In summary, we systematically derived a scheme which is equivalent
to the original Dolinar receiver for implementing the minimum
error discrimination of binary coherent-state signals. This scheme
corresponds to a spatial version of the Dolinar receiver, based
upon linear optics, photodetectors, continuous measurement and
feedforward. In our approach, as opposed to previous works, we
focus on the (asymptotically) perfect implementation of a given
projection measurement. In order to derive the Dolinar-type
scheme, we consider the projection measurement that corresponds to
the minimum error discrimination of binary coherent-state signals.
The derivation then relies on the constraint that, for
discriminating the orthogonal states of the measurement basis, the
conditional states in each detection and feedforward step must
remain orthogonal. This new derivation method does not require
complicated optimization procedures and is applicable to various
kinds of projection measurements. We showed that not only the
particular measurement associated with the coherent-state signals
treated in Dolinar's original proposal, but also other types of
projection measurements can be implemented via continuous
measurement and feedforward. Significantly, in our approach,
optimal measurement performance, i.e., the maximum measurement
efficiency allowed by quantum mechanics, is achieved without the
need of expensive entangled auxiliary resources. However, finite
feedforward plus arbitrarily many auxiliary photons must be
replaced by arbitrarily fast feedforward based on arbitrarily weak
measurements.

Although we considered only projection measurements of a
single-mode field in this paper, our methodology might be
applicable to more general scenarios as well, including
generalized measurements or joint projective measurements for more
modes. While the possibility of a linear-optics implementation of
such scenarios has been studied already
\cite{vanLoock03,Calsamiglia02}, it remains an open question
whether the approach based on continuous measurement and
feedforward allows for an optimal efficiency also in these more
general schemes.

Finally, an important question is whether our scheme can be used
in a real, necessarily finite, physical implementation. In order
to address this question, one would have to consider a discrete
analogue of our scheme, including finite feedforward and weak, but
finite measurements. One may then determine bounds in terms of
suitable figures of merit, e.g. the minimum average error or the
maximum success probability, for given feedforward resources.

\begin{acknowledgements}
The authors thank M.~Ban for helpful discussions. 
This work was supported by the DFG under the Emmy-Noether
program, the EU FET network RAMBOQ and the network of competence QIP of
the state of Bavaria.
\end{acknowledgements}


\begin{references}
\bibitem{energyloss}
Regardless of the number of photodetections, the signal loses its
energy during the measurement process. If the signal contains a
vacuum-state component, a zero-photon counting event increases the
probability amplitude for the vacuum [the second exponential term
in Eq.~(\ref{eq:S_L1-L0})]. The average energy of the signal then
decreases. Even if the initial signal does not contain the vacuum
component, e.g. a single-photon state, the process in our setup
still acts as a dissipative process. Since a displacement is not
an energy preserving operation, it redistributes the amplitude
coefficients of the partial state [the mode B in
Fig.~\ref{fig:ContinuousMeasurement}(a)] across the different
number states including the vacuum state. As a result, even if the
photon counting projects the partial state onto the vacuum, the
partial state may have contained non-zero energy before the
displacement. In other words, this event can create some
vacuum-state component in the remaining signal [mathematically,
this is due to the annihilation operator in the first exponential
term of Eq.~(\ref{eq:S_L1-L0})]. As mentioned above, once the
state includes the vacuum component, the state loses its energy
due to photon counting.
\bibitem{finiteN_tot}
Since the average energy of each signal is finite and
the displacement function in Eq.~(\ref{eq:LO})
is square integrable, $N_{\rm tot}$ does not diverge
even in the limit of $L \to \infty$.
\end{references}
\end{document}